\documentclass[12pt,a4paper,english]{article}
\usepackage{palatino}
\usepackage[T1]{fontenc}
\usepackage[latin1]{inputenc}
\usepackage{subfigure}
\usepackage{graphicx}

\makeatletter

\newlength{\lp}
\usepackage{amsfonts}

\usepackage{babel}
\makeatother
\begin{document}

\title{Flatness in Disc Galaxy Dynamics sans MOND sans Dark Matter}

\maketitle
\begin{abstract}
In previous work we have established that a particular quasi-classical
theoretical viewpoint about the nature of gravitation in galaxy discs
provides extremely high concordance (comparable with MOND) with the
observed dynamics in optical discs. 

In this paper, we consider the possibility that the optical disc and
the flat radio disc represent \emph{distinct} dynamical regions which
are inherently described by distinct phases of the same over-arching
theory of which the quasi-classical theory is but one phase. This
is a defining moment for the theory under discussion since the only
natural generalization available is that from the existing quasi-classical
(one-clock) model into a quasi-relativistic (two-clock) model. The
quasi-relativistic disc turns out to be a stationary \emph{hyperbolic}
disc within which information propagates along characteristics and
within which the rotation curve is necessarily \emph{exactly} flat. 

The transition between the two components of the disc is managed by
physically determined jump conditions which are represented by a set
of non-linear algebraic equations. These equations predict the existence
of a \emph{forbidden region} within the parameter space which coincides
exactly with that part of the parameter space in which there is an
already observed highly significant underpopulation of galaxy discs.
\end{abstract}

\section{Introduction}

\subsection{A globally inertial material vacuum}

This is the third paper in a series which has as its purpose a reformulation
of gravitation theory on the basis of what might be called an {}`ultra-strong
Machian worldview'. The initial paper (Roscoe \cite{key-32}, astro-ph/0107397)
asked the question: 

\begin{quote}
\emph{Is it possible to conceive a globally inertial space \& time
with is irreducibly associated with a non-trivial global mass distribution
and, if so, what are the properties of this distribution?}
\end{quote}
This question received a positive answer subject to the conditions
that:

\begin{itemize}
\item the global mass distribution concerned is fractal, $D=2$; 
\item the constituent particles of this distribution are photon-like in
the sense that they have a velocity property, $\mathbf{v}$, associated
with them satisfying $|\mathbf{v}|=const$;
\item the velocity property is not a true velocity but is, rather, a conversion
factor between times scales and length scales exactly like Bondi's
interpretation of light velocity, $c$. 
\end{itemize}
For obvious reasons, the foregoing was considered to be a description
of a rudimentary material vacuum.

\subsection{Gravitation as perturbations in the material vacuum}

The second paper (Roscoe \cite{key-12}, astro-ph/0306228) was concerned
with the question of how gravitational processes could arise in this
material vacuum. It began by showing how pointwise perturbations of
the $D=2$ distribution gave rise to classical Newtonian gravitational
processes for test particles placed in the region of the perturbation
and went on to derive the equations of motion for an idealized spiral
galaxy. This theory was then applied to model the rotation curves
for a small sample of low surface brightness (LSB) galaxies giving
results that were directly comparable in quality to those obtained
from the MOND algorithm - and far superior to anything achieved by
the multi-parameter dark matter models.

\subsection{Flat rotation curves and an unsought-for bonus}

However, whilst the theory of the second paper, which we shall refer
to as the quasi-classical (one-clock) theory, works extremely well
within the optical disc, it is not able to generate the flat extensions
to rotation curves usually associated with the radio disc. The purpose
of this third paper is to show how a natural generalization of the
quasi-classical one-clock theory into a quasi-relativistic%
\footnote{By quasi-relativistic, we mean a formulation in which the invariant
line element has the general structure $ds^{2}\equiv f(x,y,z,t,dx,dy,dz,dt)$
but within which $ds$ is \emph{not} defined as an element of proper
time. This latter quantity arises in an entirely natural, but quite
different, way.%
} two-clock theory solves this problem in a direct and simple way.
In essence, we find that the rotation curve of the quasi-relativistic
disc is necessarily \emph{everywhere flat.} Thus, in the presented
worldview, galaxy discs exist in two phases - one, the optical disc,
which is described by the quasi-classical theory and the other, the
radio disc, described by the quasi-relativistic theory. The transition
from the quasi-classical disc to the quasi-relativistic disc, which
can be considered analogous to the transition between a subsonic and
a supersonic flow in a compressible fluid, is managed by a set of
physically derived jump conditions which are represented by a set
of non-linear algebraic equations. 

As an unsought-for bonus, it transpires that the jump-condition equations
become degenerate in exactly that part of the disc galaxy parameter
space which we already know to be very significantly underpopulated
in such objects in the four large samples that we have analyzed (Roscoe
\cite{key-16}). In other words, the theory predicts the difficulty
of forming galaxy discs in exactly that part of the parameter space
which is already known to be significantly underpopulated by disc
galaxies.

\section{Why bother? Some difficulties faced by the standard model}

Although not yet widely accepted as problematical, it is a fact that
there exist various well-founded observations and difficulties which,
if taken at their face value, would cast reasonable doubt about the
long-term viability of the standard cosmological model. These fall
into two broad categories which we can classify as \emph{redshift
phenomena} and as \emph{non-redshift phenomena.} The point can be
illustrated by reference to just two examples which can be considered
to serve as 'straws in the wind'.

For more than two decades, Tifft (see \cite{key-51} for an early
reference) has been making claims about the existence of what he terms
as \emph{redshift periodicities.} The claims have become increasingly
elaborate over the years but, at their most simple, they relate to
the supposed existence of a periodicity at around $72$km/sec between
galaxies in clusters and one at around $36$km/sec between our own
galaxy and galaxies in the local group, out to about $2000$km/sec.
Unfortunately, Tifft was never able to support his hypotheses with
statistical analyzes rigorous enough to match the extreme nature of
the claims. However, sometime in the early 1990s Napier brought scepticism
and a long experience in dealing with noisy periodic data in the geophysical
and cometary record (variously \cite{key-52,key-54,key-55,key-56})
to the problem . There followed a short series of little-known papers
by Napier \& Guthrie (\cite{key-57,key-59,key-60}). These papers
were remarkable on two counts: firstly, for their statistical rigour
and secondly for their conclusion - that the Tifft hypotheses relating
to the $72$km/sec and the $36$km/sec periodicities were supported
on the (non-Tifft) samples used by these authors at the level of virtual
statistical certainty (see specifically \cite{key-60}). The basic
technical problems associated with these analyzes revolve around the
difficulties involved in finding line-profiles that are symmetric
enough to allow HI determinations of galaxy redshifts to within better
than $5$km/sec - it can be shown that anything much worse than this
makes it impossible to reliably detect any periodicities at the $36$km/sec
level which might exist. 

So far as \emph{non-redshift phenomena} are concerned, we need only
cite the problems associated with modelling rotation curves in disc
galaxies in general - and in low surface brightness galaxies in particular
- using the multi-parameter dark matter models, and then compare this
with the simplicity and effectiveness with which the one-parameter
MOND recipe (see Milgrom \cite{key-61,key-62,key-63} for early predictions)
succeeds in this task. Of course, MOND is not the answer for it is
not even a theory - but the details of its successes over a wide variety
of applications (see variously McGaugh, de Blok and Sanders \cite{key-65,key-66,key-67})
should be raising serious questions about the standard model's failures
in this respect - notwithstanding well-known problems experienced
with MOND when modelling galaxy clusters.

\subsection{An historic perspective on the proposed alternative view}

Given that the future of the standard model is by no means secure,
then there is room to consider alternatives, such as the one being
considered here. This alternative view was originally driven by what
could be termed as an 'ultra-strong Machian worldview' and this remains
as the core philosophy. But, interestingly, the theory which has emerged
- which employs all the machinery more usually associated with curved
spacetime theories - falls directly into the category of \emph{LeSage-}type
theories. See the introduction of Roscoe \cite{key-12} (astro-ph/0306228)
for a brief discussion of this.

LeSage \cite{key-49} was interested in the causal mechanisms which
he believed must underly Newtonian gravitation theory and he proposed
the existence of an isotropic and homogeneous substratum (in modern
terms, a material vacuum) consisting of microscopic particles in states
of agitated motion. These particles were envisaged as colliding with,
and transferring momentum to, ordinary material bodies. For one body
alone in this agitated substratum, the overall effect of the transferred
momentum was seen to be neutral. However, when two such ordinary bodies
were near each other, then they provided mutual 'shade' to each other
from the bombardment of the substratum particles thereby creating
an asymmetric imbalance in the rate of momentum transfer from the
substratum to the two bodies - the net effect being that the two bodies
experienced net forces which would appear as mutual attraction. When
the details of schemes such as this are worked through, it is a simple
matter to recover a quantitatively correct 'Newtonian gravitational
force'.

The LeSage theory, and many others of a similar kind that have appeared
over the years (see for example, Edwards \cite{key-50} for a good
review) envisage the substratum particles in a deeply classical way
- that is, as essentially microscopic versions of macroscopic classical
bodies existing within a pre-determined inertial space \& time. By
contrast, the 'substratum particles' associated with the theory of
the inertial frame developed by Roscoe \cite{key-32} (astro-ph/0107397)
are the very antithesis of the classical momentum-bearing particle:
specifically, they do not exist \emph{within} space \& time - rather,
\emph{}what we call space \& time can be considered as a \emph{metaphor}
for the relationships which exist between these vacuum particles -
in effect, they \emph{define} what we call space \& time. Thus, according
to this theory, a globally inertial space \& time is irreducibly associated
with a material vacuum within which the vacuum particles are fractally
distributed with $D=2$. Conventional gravitational processes arise
in a given locality when this distribution is perturbed in that locality.
Consequently, an intrinsic assumption of our approach is that conventional
bodies act as sinks for the vacuum particles and thereby act as local
perturbers of the $D=2$ equilibrium distribution.

\section{Review of the middle-disc data analysis}

\subsection{Disc galaxy rotation curves}

When one studies rotation curves {}`in the whole' one is typically
confronted by various complexities which would appear to defy any
chance of a simple description. These vary from non-monotonic behaviour
(typically) near the central regions, through \emph{generally} monotonically
rising sections through, finally, abrupt transitions into almost perfect
flatness somewhere in the boundary between the optical and the radio
parts of the disc. Astronomers have, traditionally, fitted very complicated
functional forms to rotation curves in attempts to provide global
classifications of their shapes. (cf Courteau \cite{key-46}, Rix
et al \cite{key-45}). In this paper, we describe the second step
of a quite different strategy whereby we consider the possibility
that galaxy discs might reasonably be partitioned into three distinct
dynamical regions, each to be considered separately within the context
of the same general theory and merged into a whole at their joining
boundaries by appropriate jump conditions:

\begin{itemize}
\item an innermost region dominated by the central bulge where the dynamics
frequently appear messy and complicated;
\item a mid-region comprising most of the optical disc where the dynamics
appear to be reasonably regular; 
\item an outer region comprising the radio disc where the dynamics gives
rise to the very peculiar constancy of rotational velocity.
\end{itemize}
The phenomenology of the optical mid-regions has already been comprehensively
analyzed in Roscoe \cite{key-42,key-16} whilst the related theory
has been developed in Roscoe \cite{key-32,key-12} (astro-ph/0107397,
astro-ph/0306228). Correspondingly, our main task here is to show
how the flat radio disc can be understood from the same general theoretical
perspective and finally merged seamlessly with the optical mid-region.
The process by which this is done also makes clear how the overall
task, that of including the \emph{three} dynamical regions within
a unified theoretical perspective, can finally be completed.

\subsection{General comments}

The mid-1990's was an optimal time for our partitioning approach since
Persic \& Salucci \cite{key-34} had just made available a sample
of 900 folded optical rotation curves (ORCs) which originated from
the Southern Sky survey of Mathewson et al \cite{key-18}. This sample
was notable for three things: 

\begin{itemize}
\item being purely optical it contained no radio disc component;
\item it was sufficiently large that it could support a rigorous statistical
analysis;
\item the individual ORCs were very accurately folded - this quality being
driven by Persic \& Salucci's intention of using them for studying
the interior dynamics of individual discs. 
\end{itemize}
The availability of this large high-quality sample meant that there
was a reasonable chance of gaining a genuine insight into the generic
behaviour of dynamics in the middle-disc regions of galaxies. Furthermore,
the quality of Persic \& Salucci's folding process turned out to be
extremely important for it added crucial precision to the statistical
study we had in mind.

\subsection{Summary of previous results}

The most simple solution provided by the quasi-classical theory when
applied to modelling an \emph{idealized} disc stated that rotation
velocities followed power laws $V=AR^{\alpha}$ where $(A,\alpha)$
are parameters that define individual idealized discs. This solution
turns out to be merely the most simple in a class of much more complex
solutions, but it defined the basic question which has driven most
of this author's data analyzes: specifically, \emph{to what extent
can rotation velocities in real galaxy discs be said to be described
by simple power laws?} This question turned out to be an immensely
productive one. The results arising from pursuing it - both directly
and indirectly - are listed below.

\begin{itemize}
\item Over large samples of late-type galaxies ($Sb..Sd$), and to an extremely
high \emph{}statistical precision, rotation velocities in the \emph{outer
regions} of optical discs can be represented by power-laws of the
type\[
V=AR^{\alpha}\rightarrow\frac{V}{V_{0}}=\left(\frac{R}{R_{0}}\right)^{\alpha}\rightarrow\ln A=\ln V_{0}-\alpha\ln R_{0}\]
 where the scaling constants, $(V_{0},R_{0})$ are very strong functions
of galaxy luminosity properties (Roscoe \cite{key-42,key-16});
\item The power-law structure in the outer regions of optical discs gives
automatic rise to the classical Tully-Fisher relations which are properly
calibrated for $I$-band luminosity data and $R$-band luminosity
data according to whether the sample being analyzed uses $I$-band
or $R$-band photometry (Roscoe \cite{key-16});
\item The pursuit of the power-law question led directly to the discovery
of the \emph{discrete dynamical states} phenomenology in disc galaxies
according to which the distribution of the parameter $A$ in the law
$V=AR^{\alpha}$ is multi-peaked with four strong peaks satisfying
$\ln A\approx$$3.90$, $4.17$, $4.72$ and $5.11$ being identified.
This translates into the statement that rotational velocities at $1$kpc
tend to cluster around the values of $49$km/sec, $65$km/sec, $112$km/sec
and $166$km/sec (Roscoe \cite{key-33,key-16});
\item The statistical success of the simple power-law formulation for outer-disc
optical rotation curves prompted a much deeper study of the underlying
quasi-classical theory. Subsequently, we were able to show that it
could be applied to model the dynamics of low surface brightness galaxies
(LSBs) giving results that hitherto have only been matched by MOND
(Roscoe \cite{key-12}, astro-ph/0306228).
\end{itemize}
It is the general success of the quasi-classical model, as outlined
above, which has encouraged the development of the work which forms
the main body of this paper.

\subsection{Summary of new results }

The current paper provides three additional significant results which
can be stated as:

\begin{itemize}
\item A fresh, but simple, insight into the quasi-classical model provides
a qualitative understanding of why the power-law model $V=AR^{\alpha}$
fits the generality of middle-disc dynamics as well as it does - even
though it is a simple matter to find examples of individual discs
which are in clear departure from any power-law description.
\item the rotation curve flatness problem is resolved within the theory:
specifically, the flat exterior sections of rotation curves arise
as the \emph{only} solutions of the two-clock quasi-relativistic generalization
of the one-clock quasi-classical theory;
\item the jump conditions between the quasi-classical and the quasi-relativistic
parts of the disc form a set of non-linear algebraic equations which
become degenerate in that part of the parameter space which corresponds
\emph{exactly} to the deep valley between the $\ln A=4.17$ and the
$\ln A=4.72$ peaks of the $\ln A$ frequency diagram. That is, the
jump conditions predict the existence of a forbidden zone in the parameter
space which appears to exist on the data.
\end{itemize}

\section{Review of the quasi-classical disc theory\label{sec:Review-of-Time-Independent}}

A brief review of the theory underlying the quasi-classical middle-disc
model (Roscoe \cite{key-32,key-12} astro-ph/0107397, astro-ph/0306228)
provides a useful precursor for the main analysis of this paper and
provides the first new result.

\subsection{Physical time and mathematical closure\label{sub:Physical-time-and} }

The basic relation of the quasi-classical theory is given by

\begin{equation}
g_{ab}\equiv\nabla_{a}\nabla_{b}{\mathcal{M}}\equiv{\frac{\partial^{2}{\mathcal{M}}}{\partial x^{a}\partial x^{b}}}-\Gamma_{ab}^{k}{\frac{\partial{\mathcal{M}}}{\partial x^{k}}},\label{eq0}\end{equation}
where the indices take values $1..3$ and ${\cal M}$ provides a scalar
representation of the mass distribution concerned - but see Roscoe
\cite{key-12} (astro-ph/0306228) for a complete discussion. In the
present case, we are concerned with modelling disc galaxies and here
${\cal M}(r)$ has the straightforward interpretation as the amount
of mass contained within a radius $r$ in a disc of unit thickness.
Furthermore, for disc galaxies, the assumption of cylindrical symmetry
is appropriate implying $\Gamma_{ab}^{k}\equiv0$. Using cylindrical
polars, we then find that the line element can be written as\[
ds^{2}=g_{ij}dx^{i}dx^{j}=\frac{\partial^{2}{\cal M}}{\partial r^{2}}dr^{2}+\left(\frac{1}{r}\frac{\partial{\cal M}}{\partial r}\right)r^{2}d\theta^{2}.\]
Introducing the quantities $R=r/r_{0}$ and $T=t/t_{0}$ where $(r_{0},t_{0})$
are scaling constants, we get the dimensionless form\begin{equation}
ds^{2}=g_{ij}dx^{i}dx^{j}=\frac{\partial^{2}{\cal M}}{\partial R^{2}}dR^{2}+\left(\frac{1}{R}\frac{\partial{\cal M}}{\partial R}\right)R^{2}d\theta^{2}\equiv A\, dR^{2}+B\, R^{2}d\theta^{2}.\label{eq-3}\end{equation}
The equations of motion can now be written down in the usual way using
the variational principle\[
{\cal I}=\int\sqrt{g_{ij}\dot{x}^{i}\dot{x}^{j}}d\tau.\]
However, because ${\cal I}$ is homogeneous degree zero in the temporal
ordering parameter, $\tau$, then the resulting equations of motion
are invariant wrt arbitrary transformations of this parameter. This
has two related consequences: firstly, it implies that $\tau$ does
not, in general, represent physical time and, secondly, the resulting
equations of motion form an incomplete set. In fact, because the analysis
is two-dimensional, the variational principal only gives one independent
equation of motion, namely\[
\frac{d}{d\tau}\left\{ R^{2}\dot{\theta}\frac{B}{{\cal L}}\right\} =0\rightarrow R^{2}\dot{\theta}B=\sqrt{m_{1}}\,\left(A\dot{R}^{2}+BR^{2}\dot{\theta}^{2}\right)^{1/2}\]
which corresponds to the conservation of angular momentum in purely
classical theory. The situation is, of course, identical for General
Relativity - there, the equations of motion arising from the variational
principle are completed by introducing the concept of particle proper
time. In the case under discussion (see Roscoe \cite{key-12} or astro-ph/0306228
for a complete discussion) this was not an option, and a different
solution had to be sought: Certain symmetry arguments led to the conclusions
that, within the context of the application of the theory to disc
galaxies, 

\begin{itemize}
\item the forces acting within the disc are not necessarily central;
\item the ratio of the magnitudes of the transverse forces to the radial
forces along any radius within the disc is constant.
\end{itemize}
This led to the closure of the system in the following way: the line
element is given by\begin{equation}
ds^{2}=A\, dR^{2}+B\, R^{2}d\theta^{2}\label{eq1A}\end{equation}
where $A$ and $B$ are defined in (\ref{eq-3}). Since the closure
information is derived from the ratio of force components, it is necessary
to define what we mean by \emph{force} in a disc with line element
given by (\ref{eq1A}). To this end, we factorize this latter relation
as\begin{equation}
ds^{2}\equiv d\mathbf{s}\cdot d\mathbf{s},\,\,\,\, d\mathbf{s}\equiv S_{1}\sqrt{A}\, dR\,\mathbf{\hat{R}}+S_{2}\sqrt{B}\, R\, d\theta\,\hat{\theta}\label{eq1F}\end{equation}
where $S_{1}=\pm1,\, S_{2}=\pm1$ and where, in the standard notation,
${\bf \mathbf{\hat{R}}}=(\cos\theta,\sin\theta)^{T}$ and $\hat{\theta}=(-\sin\theta,\cos\theta)^{T}$.
Hence, introducing an arbitrary temporal ordering parameter, $\tau$,
we can define a generalized velocity as\[
\mathbf{V}=S_{1}\sqrt{A}\,\dot{R}\,\mathbf{\hat{R}}+S_{2}\sqrt{B}\, R\dot{\theta}\,\hat{\theta}\]
from which a generalized acceleration vector can be derived as\[
\mathbf{\dot{V}}=\left\{ S_{1}\frac{d}{d\tau}\left(\sqrt{A}\,\dot{R}\right)-S_{2}\sqrt{B}\, R\,\dot{\theta}^{2}\right\} \mathbf{\,\hat{R}}+\left\{ S_{2}\frac{d}{d\tau}\left(\sqrt{B}R\dot{\theta}\right)+S_{1}\sqrt{A}\,\dot{R}\,\dot{\theta}\right\} \,\hat{\theta}.\]
However, we also require that the ratio of the magnitude of the transverse
force component to the radial force component is to be constant, so
that we get the additional equation\begin{equation}
\left\{ S_{1}\frac{d}{d\tau}\left(\sqrt{A}\,\dot{R}\right)-S_{2}\sqrt{B}\, R\,\dot{\theta}^{2}\right\} =K\left\{ S_{2}\frac{d}{d\tau}\left(\sqrt{B}R\dot{\theta}\right)+S_{1}\sqrt{A}\,\dot{R}\,\dot{\theta}\right\} \label{eq1C}\end{equation}
for some constant $K$. Since this equation is \emph{not} invariant
under arbitrary transformations of the parameter $\tau$ then, effectively,
it selects a specific $\tau$ from all the possibilities and defines
it as \emph{physical time} - that is, as that definition of time which
is consistent with our constraint on disc forces. 

Thus, at face value, we now have two equations for the system - one
from the variational principle (not shown) and (\ref{eq1C}), above.
However, it transpires that if one assumes that light traces mass
in a disc galaxy (that is, no significant dark matter) then, typically
for light-inferred mass distributions, one finds\[
A\equiv\frac{\partial^{2}{\cal M}}{\partial R^{2}}<0\]
which corresponds to density distributions which fall off more quickly
than $1/R$ but no quicker than $1/R^{2}$. Since $B$, defined at
(\ref{eq-3}), is necessarily positive for \emph{any} ${\cal M}$,
then equation (\ref{eq1C}) is necessarily complex which, in turn,
implies $K\equiv q+pj$ for real parameters $(p,q)$. Thus, in the
final analysis we end up with \emph{three} equations - which implies
that the theory determines the dynamics \emph{and} the mass distribution
for a disc galaxy.

\subsection{The power-law as a limiting velocity distribution\label{sub:The-power-law-as}}

The same theory that gave almost perfect modelling for the sample
of eight LSBs considered in Roscoe \cite{key-12} (astro-ph/0306228)
(data provided by McGaugh%
\footnote{Department of Astronomy, University of Maryland, USA%
}) also provides an understanding of why the simple power-law model
provides such a high-fidelity statistical resolution of middle-disc
ORC data over large data sets. 

In effect, we find that the power-law solution, $V=AR^{\alpha}$,
is the only solution which is bounded over finite discs and that all
other possible solutions become rapidly unbounded and therefore, by
definition, cannot persist over \emph{substantial} radial ranges within
stable discs. However, according to the theory, these non-power-law
solutions \emph{can} persist over small radial ranges and are therefore,
in terms of the theory, responsible for much of the complexity commonly
observed within optical discs. However, since, by definition, stable
discs have bounded velocity distributions, then it is the power-law
which is the dominant underlying form.

To obtain a quantitative understanding of these remarks, we must consider
the basic theory: the density-distribution equation derived in Roscoe
\cite{key-12} (astro-ph/0306228) can be written as

\begin{eqnarray}
\left(\Psi^{2}-1\right)\epsilon^{2}+\frac{2}{q}\left\{ p(1+\Psi^{2})+(1+p^{2}+q^{2})\Psi\right\} -\left(\Psi^{2}-1\right) & = & 0\nonumber \\
\label{eq20}\\{\rm where}\,\,\,\epsilon^{2}\equiv-1-\frac{R}{\rho}\frac{d\rho}{dR}\,\,\,{\rm and}\,\,\,\Psi^{2}\equiv1-\frac{2\pi\rho}{m_{1}}R^{2},\,\,\,\rho\equiv\frac{1}{2\pi R}\frac{\partial{\cal M}}{\partial R}\nonumber \end{eqnarray}
where $\rho$ is surface density, ${\cal M}$ is the basic mass function
of the theory, $m_{1}$ is a scaling parameter having dimensions of
mass and $(p,q)$ are free dimensionless parameters. The corresponding
rotational velocity equation can be written as\begin{equation}
\frac{1}{V}\frac{dV}{dR}=\frac{1}{R}\left[\frac{1}{2}\left(1+\epsilon^{2}\right)-\frac{p}{q}\epsilon-\left(\frac{p^{2}+q^{2}}{q}\right)\frac{\epsilon}{\Psi}\right].\label{eq21}\end{equation}

It is easily shown that if\begin{equation}
{\cal M}=m_{0}\ln d_{0}R\label{eq20A}\end{equation}
 where $m_{0}$ has dimensions of mass and $d_{0}$ is dimensionless,
then $\rho\sim1/R^{2}$, $\epsilon^{2}=1$ and $\Psi=const$ etc with
the final consequence that the velocity equation gives $V=AR^{\alpha}$
exactly. But also, by the considerations of \S\ref{sub:Physical-time-and},
we know that $\rho$ cannot fall of \emph{more} rapidly than $1/R^{2}$
- otherwise equation (\ref{eq1C}) becomes real, and the whole analysis
becomes invalidated. We are therefore left to consider solutions for
which $\rho\sim1/R^{k},\,\, k<2$: from (\ref{eq20}) it is easy to
see that, for such solutions, $\Psi^{2}\rightarrow0$ as $R$ increases.
But, by (\ref{eq21}), $dV/dR\rightarrow\infty$ as $\Psi\rightarrow0$
so that, as stated, all other possible solutions are unbounded over
any finite disc. 

To conclude, the power-law solution is the only solution which is
bounded over a finite disc and therefore, according to the theory,
provides the underlying dynamical {}`shape' of any stable disc.

\section{The Development of the quasi-relativistic disc-model}

We show, in appendix \ref{sec:The-possibility-of}, that it is impossible
to have non-trivial power-law solutions merging into flat solutions
within the context of the quasi-classical one-clock theory. The solution
to this problem turns out to lie in the quasi-relativistic generalization
of the quasi-classical model: Specifically, it transpires that the
flat component of rotation curves is defined in a hyperbolic space
which arises directly from the quasi-relativistic generalization whilst
the optical mid-region component remains associated with the quasi-classical
model. The boundary between the two components can be considered analogous
to a shock front, with the transition across this boundary being managed
by appropriately defined jump conditions.

\subsection{The Euler-Lagrange component}

The basic definition for the metric tensor, given at (\ref{eq0})
remains unchanged - except that now, the indices vary over $1..4$
where $x^{4}\equiv ct$ and $t$ represents conventional clock time
and $c$ is a scaling parameter with the dimensions of velocity. In
cylindrical polars, the line element becomes\[
ds^{2}=g_{ij}dx^{i}dx^{j}=\frac{\partial^{2}{\cal M}}{\partial r^{2}}dr^{2}+\left(\frac{1}{r}\frac{\partial{\cal M}}{\partial r}\right)r^{2}d\theta^{2}+2\frac{\partial^{2}{\cal M}}{\partial r\partial t}drdt+\frac{\partial^{2}{\cal M}}{\partial t^{2}}dt^{2}\]
which, in dimensionless form, is given by\[
ds^{2}=g_{ij}dx^{i}dx^{j}=\frac{\partial^{2}{\cal M}}{\partial R^{2}}dR^{2}+\left(\frac{1}{R}\frac{\partial{\cal M}}{\partial R}\right)R^{2}d\theta^{2}+2\frac{\partial^{2}{\cal M}}{\partial R\partial T}dRdT+\frac{\partial^{2}{\cal M}}{\partial T^{2}}dT^{2}\]
where $R=r/r_{0}$ and $T=t/t_{0}$ where $(r_{0},t_{0})$ are scaling
constants, as before. Now introduce\begin{equation}
A\equiv\frac{\partial^{2}{\cal M}}{\partial R^{2}},\,\,\, B\equiv\frac{1}{R}\frac{\partial{\cal M}}{\partial R},\,\,\, D\equiv2\frac{\partial^{2}{\cal M}}{\partial R\partial T},\,\,\, E\equiv\frac{\partial^{2}{\cal M}}{\partial T^{2}},\label{eq1D}\end{equation}
noting that $(A,B,C,D)$ all have dimensions of mass, so that\begin{equation}
ds^{2}=A\, dR^{2}+B\, R^{2}d\theta^{2}+D\, dR\, dT+E\, dT^{2}\label{eq1B}\end{equation}
from which we get the variational principle\begin{equation}
{\cal I}=\int{\cal L}d\tau,\,\,\,{\cal L}=\sqrt{g_{ij}\dot{x}^{i}\dot{x}^{j}}\equiv\left(A\dot{R}^{2}+BR^{2}\dot{\theta}^{2}+D\dot{R}\dot{T}+E\dot{T}^{2}\right)^{1/2}.\label{eq1E}\end{equation}
As before, this is homogeneous degree zero in the parameter $\tau$,
so that the equations of motion form an incomplete set given by\begin{eqnarray}
\frac{d}{d\tau}\left\{ R^{2}\dot{\theta}\frac{B}{{\cal L}}\right\}  & = & 0\rightarrow R^{2}\dot{\theta}B=\sqrt{m_{1}}\,\left(A\dot{R}^{2}+BR^{2}\dot{\theta}^{2}+D\dot{R}\dot{T}+E\dot{T}^{2}\right)^{1/2}\label{eq5}\end{eqnarray}
where $m_{1}$ is a scaling constant with dimensions of mass, together
with\begin{equation}
\frac{d}{d\tau}\left\{ \frac{1}{2{\cal L}}\left(D\dot{R}+2E\dot{T}\right)\right\} -\frac{\partial{\cal L}}{\partial T}=0.\label{eq6}\end{equation}
In a classical context, the equation corresponding to (\ref{eq5})
would give angular momentum conservation. Clearly, angular momentum
is not conserved here (the forces are not assumed central) - but something
is. For convenience, we shall refer to (\ref{eq5}) as the equation
for the conservation of \emph{generalized angular momentum.}

\subsection{Physical time and mathematical closure\label{sub:Temporal closure}}

As before, the problem is that the system is invariant wrt to arbitrary
(monotonic) transformations of the temporal ordering parameter, $\tau$,
which implies that physical time is not defined - some physical condition
is missing from the system. To progress, we \emph{could} adopt the
GR solution of introducing the notion of particle proper time - but
that would entail ignoring the considerable success of the earlier
approach (above and Roscoe \cite{key-12}, astro-ph/0306228) which
depended critically upon the idea that the ratio of the magnitudes
of transverse to radial force components was constant within any given
disc. We therefore seek to generalize the approach of \S\ref{sub:Physical-time-and}
to the present case using the line element (\ref{eq1B}). In the following,
we sketch out the analysis, giving the details in appendix \ref{sec:Temporal closure}.
The only factorization of the line element (\ref{eq1B}) similar to
that of (\ref{eq1F}) turns out to be\begin{equation}
ds^{2}\equiv d\mathbf{s}\cdot d\mathbf{s}\,\,\,{\rm where}\,\,\, d\mathbf{s}=S_{1}\sqrt{A}\, dR\,\mathbf{\hat{R}}+S_{2}\sqrt{B}\, R\, d\theta\,\hat{\theta}+S_{3}\sqrt{E}\, dT\,\mathbf{\hat{R}},\label{eq2}\end{equation}
where $\mathbf{\hat{R}}$ and $\mathbf{\hat{\theta}}$ are unit vectors
as before, subject to the condition on ${\cal M}$ that\begin{equation}
D=2S_{1}S_{3}\sqrt{AE}\rightarrow\left(\frac{\partial^{2}{\cal M}}{\partial R\partial T}\right)^{2}=\frac{\partial^{2}{\cal M}}{\partial R^{2}}\,\frac{\partial^{2}{\cal M}}{\partial T^{2}}.\label{eq4}\end{equation}
We can now use (\ref{eq2}) to define a generalized velocity, $d\mathbf{s}/d\tau\equiv\mathbf{V}$,
and a corresponding generalized acceleration, $d\mathbf{V}/d\tau\equiv\mathbf{\dot{V}}$,
in the space in a way which is directly analogous to the development
of \S\ref{sub:Physical-time-and}. The arbitrariness associated with
the temporal ordering parameter, $\tau$, is removed by applying the
physical condition that the ratio of radial to transverse generalized
accelerations is to be constant within the disc. Adopting the conventions
that\[
\sqrt{A}=+j\sqrt{-A},\,\,\,\sqrt{E}=+j\sqrt{-E},\,\,\, S_{1}S_{2}\rightarrow S_{0},\,\,\, S_{2}S_{3}\rightarrow S_{1}\]
then this physical condition gives rise to the additional equations\begin{eqnarray}
\frac{d}{d\tau}\left(\sqrt{B}V_{\theta}\right) & = & -q\sqrt{B}\frac{V_{\theta}^{2}}{R}-p\frac{d}{d\tau}\left\{ S_{0}\sqrt{-A}V_{R}+S_{1}\sqrt{-E}\dot{T}\right\} \label{eq7}\\
\nonumber \\\frac{V_{\theta}}{R}\left(S_{0}\sqrt{-A}V_{R}+S_{1}\sqrt{-E}\dot{T}\right) & = & -p\sqrt{B}\frac{V_{\theta}^{2}}{R}+q\frac{d}{d\tau}\left\{ S_{0}\sqrt{-A}V_{R}+S_{1}\sqrt{-E}\dot{T}\right\} \label{eq8}\end{eqnarray}
whilst (\ref{eq4}) becomes\begin{equation}
D=-2S_{0}S_{1}\sqrt{-A}\sqrt{-E}\rightarrow\left(\frac{\partial^{2}{\cal M}}{\partial R\partial T}\right)^{2}=\frac{\partial^{2}{\cal M}}{\partial R^{2}}\,\frac{\partial^{2}{\cal M}}{\partial T^{2}}.\label{eq9}\end{equation}
In the above, $V_{\theta}\equiv R\dot{\theta}$ and $V_{R}\equiv\dot{R}$
are the rotational and radial components of the generalized velocity,
$\mathbf{V}$, and $(p,q)$ are dimensionless parameters directly
analogous to those of \S\ref{sub:Physical-time-and}.

To summarize, our equations of motion are given by (\ref{eq5}), (\ref{eq6}),
(\ref{eq7}) and (\ref{eq8}) together with the condition (\ref{eq9})
which must be also satisfied. Thus, we now have five equations to
be solved for four unknowns, $(\dot{R},\, R\dot{\theta},\,\dot{T},\,{\cal M})$
so that the system appears to be overdetermined - unless there is
redundancy somewhere. We shall show that this is, in fact, the case.

\subsection{Linear dependencies with the system\label{sub:A-Linear-Dependancy}}

We show, in appendix \ref{sec:A-Rearrangement-of}, how (\ref{eq5}),
(\ref{eq7}) and (\ref{eq8}) lead to the equations\begin{eqnarray}
\sqrt{B}\Psi v_{\theta}-S_{0}\sqrt{-A}v_{R} & = & S_{1}\sqrt{-E}\label{eq11}\\
\nonumber \\\left\{ q+\frac{(1+p\Psi)(p+\Psi)}{q\Psi}\right\} v_{\theta}-\frac{R}{\Psi}\frac{\partial\Psi}{\partial R}v_{R} & = & \frac{R}{\Psi}\frac{\partial\Psi}{\partial T}\label{eq12}\end{eqnarray}

where\begin{equation}
v_{\theta}\equiv\frac{V_{\theta}}{\dot{T}},\,\,\, v_{R}=\frac{V_{R}}{\dot{T}},\,\,\,\Psi\equiv S_{2}\sqrt{1-\frac{BR^{2}}{m_{1}}},\,\,\, S_{2}=\pm1\label{eq9A}\end{equation}
 The structure of (\ref{eq11}) and (\ref{eq12}) is fundamental to
the system. By removing the time-dependent terms, it is easily seen
that the quasi-classical form of this latter pair is given by\begin{eqnarray*}
\sqrt{B}\Psi v_{\theta}-S_{0}\sqrt{-A}v_{R} & = & 0\\
\\\left\{ q+\frac{(1+p\Psi)(p+\Psi)}{q\Psi}\right\} v_{\theta}-\frac{R}{\Psi}\frac{\partial\Psi}{\partial R}v_{R} & = & 0\end{eqnarray*}
from which it is immediate that non-trivial solutions for $v_{\theta}$
and $v_{R}$ can only exist if \[
\sqrt{B}R\frac{\partial\Psi}{\partial R}-S_{0}\sqrt{-A}\left\{ q+\frac{(1+p\Psi)(p+\Psi)}{q\Psi}\right\} =0.\]
This equation is, in fact, a rearranged form of the quasi-classical
mass-equation, (\ref{eq20}), which immediately suggests the possibility
that (\ref{eq11}) and (\ref{eq12}) are similarly linearly dependent.
If so, we must necessarily have the conditions\begin{eqnarray}
\sqrt{B}R\frac{\partial\Psi}{\partial R}-S_{0}\sqrt{-A}\left\{ q+\frac{(1+p\Psi)(p+\Psi)}{q\Psi}\right\}  & = & 0\label{eq14}\\
\nonumber \\\sqrt{B}R\frac{\partial\Psi}{\partial T}-S_{1}\sqrt{-E}\left\{ q+\frac{(1+p\Psi)(p+\Psi)}{q\Psi}\right\}  & = & 0\nonumber \end{eqnarray}
simultaneously satisfied. We immediately see that this is only possible
if the additional mass condition \begin{equation}
S_{1}\sqrt{-E}\frac{\partial\Psi}{\partial R}=S_{0}\sqrt{-A}\frac{\partial\Psi}{\partial T}\label{eq13}\end{equation}
holds - remember that $A,B,D,E,\Psi$ are all defined in terms of
${\cal M}$. But ${\cal M}$ must also satisfy (\ref{eq9}), so we
must now consider whether or not such a thing is possible.

\subsection{The Mass Equation\label{sub:The-Mass-Equation}}

Equation (\ref{eq9}) has two basic classes of solution, given by
a logarithmic class\begin{equation}
{\cal M}(R,T)=m_{0}\ln(1+R+T)+d_{0},\label{eq3}\end{equation}
where $(m_{0},d_{0})$ are constants, and a variables separable class.
The constant $d_{0}$ is irrelevant here since only derivatives of
${\cal M}$ appear in the theory; we ignore it from now on. The variables
separable class, which is considered in appendix \ref{sec:The-Variables-Separable}
for completeness, turns out to be associated with degeneracy in the
system and so plays no part in the main body of this paper. The logarithmic
class concerns us directly and we ask under what conditions (if any)
the solution (\ref{eq3}) can satisfy condition (\ref{eq13}). Using
the definitions (\ref{eq1D}) and (\ref{eq9A}) we easily find that
the logarithmic class can only satisfy (\ref{eq13}) if the condition\[
-S_{0}S_{1}\frac{(1+T)}{R}=1\]
is satisfied. Noting that $R=-(1+T)$ is excluded by the structure
of (\ref{eq3}), then we must have\[
R=1+T,\,\,\, S_{0}S_{1}=-1.\]
With this latter condition, we find that ${\cal M}(R,T)=m_{0}\ln(1+R+T)$
is a solution subject to $R=1+T$ and the conditions (\ref{eq14})
which now reduce to the single algebraic equation\begin{eqnarray}
Mq-S_{1}2^{5/2}\left\{ p\left(2-\frac{M}{2}\right)+S_{2}\left(1+p^{2}+q^{2}\right)\sqrt{1-\frac{M}{2}}\right\}  & = & 0\label{eq19}\end{eqnarray}
where $M\equiv m_{0}/m_{1}$, $S_{1}=\pm1$ and $S_{2}=\pm1$. To
summarize, the solution can be expressed as \begin{equation}
{\cal M}=m_{0}\ln2R\label{eq19A}\end{equation}
on the characteristic $R=1+T$ subject to the condition that the dimensionless
mass parameter $M$ satisfies (\ref{eq19}).

\section{Rotational velocity $=constant$ in the hyperbolic disc}

In this section, we obtain the extremely interesting result that,
when the mass distribution is as described in \S\ref{sub:The-Mass-Equation},
then the rotational velocity is \emph{constant} with a value determined
by the parameters $(p,q)$ and the algebraic equation (\ref{eq19}).
The four basic (but rearranged) equations of motion are:\begin{eqnarray}
\sqrt{B}\Psi V_{\theta}-S_{0}\sqrt{-A}V_{R} & = & S_{1}\sqrt{-E}\dot{T}\label{eq15}\\
\nonumber \\\frac{d}{d\tau}\left\{ \frac{1}{2{\cal L}}\left(D\dot{R}+2E\dot{T}\right)\right\} -\frac{\partial{\cal L}}{\partial T} & = & 0\label{eq16}\\
\nonumber \\\frac{d}{d\tau}\left(\sqrt{B}V_{\theta}\right) & = & -q\sqrt{B}\frac{V_{\theta}^{2}}{R}-p\frac{d}{d\tau}\left(V_{\theta}\sqrt{B}\Psi\right)\label{eq17}\\
\nonumber \\\left\{ q+\frac{(1+p\Psi)(p+\Psi)}{q\Psi}\right\} V_{\theta}-\frac{R}{\Psi}\frac{\partial\Psi}{\partial R}V_{R} & = & \frac{R}{\Psi}\frac{\partial\Psi}{\partial T}\dot{T}\label{eq18}\end{eqnarray}
The first two equations arise directly from the variational principle
(\ref{eq1E}), whilst the third and fourth equations arise from the
conditions (\ref{eq7}) and (\ref{eq8}). In \S\ref{sub:A-Linear-Dependancy}
we showed the subsidiary mass condition (\ref{eq9}) could be satisfied
by making (\ref{eq15}) and (\ref{eq18}) linearly dependent so that,
in effect, we can consider the latter of these to be redundant. Consequently,
we have three equations (\ref{eq15}), (\ref{eq16}) and (\ref{eq17})
say, for the determination of $(V_{\theta},\, V_{R},\,\dot{T)}$. 

To progress, (\ref{eq16}) must be expanded explicitly and this is
done in appendix \ref{sec:Expansion-of-an}, where we show how it
reduces to an \emph{algebraic} equation in the dimensionless velocity
$v_{\theta}$ with the solution\begin{equation}
v_{\theta}=\frac{1}{\sqrt{2-M}}\label{eq18A}\end{equation}
where $M\equiv m_{0}/m_{1}$ must satisfy (\ref{eq19}). That is,
the only possibility for rotation velocities in the disc of the quasi-relativistic
theory is that of \emph{constancy} - the rotation curve is flat.

\section{Jump Conditions}

We are concerned with the transition from the quasi-classical rising
part of the rotation curve to the flat quasi-relativistic disc and,
in all that follows, we use the superfix $*$ to denote quantities
in the quasi-relativistic disc. It transpires that everything reduces
to deriving the relationships between the quasi-classical parameters
$(m_{0},m_{1})$ and their quasi-relativistic counterparts $(m_{0}^{*},m_{1}^{*})$
and, to this end, we use jump conditions which quantify 

\begin{itemize}
\item conservation of generalized angular momentum - cf equation (\ref{eq5});
\item conservation of mass-flow across the transition boundary;
\item no-slip for the rotational flow at the transition boundary.
\end{itemize}
The first of these requires some discussion: within the context of
the present model, forces within discs are not central forces - consequently,
angular momentum is not conserved. However, if we consider (\ref{eq5})
we see that the quantity $R^{2}\dot{\theta}B/{\cal L}$ (which plays
the part here of classical angular momentum) is conserved in both
the quasi-classical and hyperbolic discs. Accordingly, it is this
quantity which must be conserved across the transition boundary.

\subsection{Conservation of Generalized Angular Momentum\label{sub:Conservation-of-Generalized}}

Reference to (\ref{eq5}) shows that the quantity $R^{2}\dot{\theta}B/{\cal L}$
must be conserved across the radial transition boundary, $R=R_{T}$
say. Using the no-slip condition $v_{\theta}^{*}=v_{\theta}$ at $R=R_{T}$
then this requirement reduces to\[
\frac{B^{*}}{{\cal L}^{*}}=\frac{B}{{\cal L}}\]
across $R=R_{T}$ . Hence, following appendix \ref{sec:Conservation-of-Generalized},
we obtain the condition for generalized angular momentum across the
transition boundary as $m_{1}^{*}=m_{1}$.

\subsection{Conservation of mass flow\label{sub:Conservation-of-mass}}

The mass-flow conservation across the transition boundary, $R=R_{T}$,
can be stated as:\[
\left.\rho^{*}v_{R}^{*}\right|_{R=R_{T}}=\left.\rho v_{R}\right|_{R=R_{T}}.\]
Hence, following appendix \ref{sec:Conservation-of-mass}, we obtain
the condition for mass conservation across the transition boundary
as\begin{equation}
\Psi=\frac{m_{0}^{*}}{m_{0}}S_{0}^{*}S_{0}\frac{\sqrt{2-M^{*}}}{2}\left(S_{2}^{*}-S_{1}^{*}\right),\,\,\, M^{*}\equiv\frac{m_{0}^{*}}{m_{1}^{*}}\label{eq27A}\end{equation}
together with\begin{equation}
\Psi\equiv S_{2}\sqrt{1-M},\,\,\, M\equiv\frac{m_{0}}{m_{1}}\label{eq27}\end{equation}
At this stage, we can note that $\Psi\neq0$ necessarily otherwise,
as is easily seen by reference to (\ref{eq21A}) and (\ref{eq21B}),
dynamics in the quasi-classical disc becomes indeterminate. It follows
immediately, from (\ref{eq27A}) above that\begin{equation}
S_{1}^{*}S_{2}^{*}=-1.\label{eq27B}\end{equation}

\subsection{Summary of jump relations}

For convenience, we collect the jump conditions in the form we use
them, below: Using (\ref{eq27B}) then (\ref{eq27A}) and (\ref{eq27})
give, after squaring,\[
\left(\frac{m_{0}^{*}}{m_{0}}\right)^{2}\left(2-M^{*}\right)=1-\left(\frac{m_{0}}{m_{1}}\right)\]
where we remember that $M^{*}\equiv m_{0}^{*}/m_{1}^{*}$. Use of
the generalized angular momentum jump condition $m_{1}^{*}=m_{1}$
now gives, after a bit of algebra,\begin{equation}
\left(\frac{m_{0}}{m_{0}^{*}}\right)^{3}M^{*}-\left(\frac{m_{0}}{m_{0}^{*}}\right)^{2}+2-M^{*}=0\label{eq28}\end{equation}
which we refer to as our first jump condition. Use of (\ref{eq27B})
in (\ref{eq27A}) gives\begin{equation}
\Psi=-\frac{m_{0}^{*}}{m_{0}}S_{0}^{*}S_{0}S_{1}^{*}\sqrt{2-M^{*}},\,\,\, M^{*}\equiv\frac{m_{0}^{*}}{m_{1}^{*}}\label{eq28A}\end{equation}
which we refer to as our second jump condition whilst our third is
simply\begin{equation}
m_{1}^{*}=m_{1}.\label{eq28B}\end{equation}

\section{Forbidden zones in the parameter space}

In the following, where the details of merging the rising part of
a rotation curve (typically, the \emph{optical} rotation curve) with
its flat radio extension are considered, we remember that $(p,q)$
are dimensionless disposable parameters which appeared firstly after
(\ref{eq1C}) via $K=q+pj$ and similarly in appendix \ref{sec:Temporal closure}.
We assume that they are fixed across the whole disc.

To simplify matters, we represent the rising part of the rotation
curve purely in terms of the power law solution $V=AR^{\alpha}$.
The theory says nothing about $A$ since this would generally be determined
by initial conditions, but is explicit about the structure of $\alpha$
in terms of the free parameters in the theory. Specifically, by (\ref{eq21})
and the sentence which follows it, we have\begin{equation}
\alpha=\left[1\mp\frac{p}{q}\mp\left(\frac{p^{2}+q^{2}}{q}\right)\frac{1}{\Psi}\right]\label{eq28C}\end{equation}
and we shall consider the constraints placed upon this quantity by
the theory. Note that, in practice, we are only concerned with $0<\alpha<1$
since there are essentially no optical rotation curves outside of
this range (Roscoe \cite{key-42,key-16}).

The constraining equations are, from (\ref{eq20}) in the case of
a power-law solution on the quasi-classical disc, the condition\[
p(1+\Psi^{2})+(1+p^{2}+q^{2})\Psi=0\]
together with \[
M^{*}q-S_{1}^{*}2^{5/2}\left\{ p\left(2-\frac{M^{*}}{2}\right)+S_{2}^{*}\left(1+p^{2}+q^{2}\right)\sqrt{1-\frac{M^{*}}{2}}\right\} =0;\,\,\, S_{1}^{*}S_{2}^{*}=-1,\]
from (\ref{eq19}) on the quasi-relativistic disc. Together with (\ref{eq28})
and (\ref{eq28A}), these represent four constraints on the five unknowns
$(p,\, q,\, M^{*},\, m_{0}/m_{0}^{*},\,\Psi)$.

\subsection{Brief computational details}

There are five unknowns and four constraints together with various
unknown signatures. A simple (inefficient with lots of redundancy!)
approach is as follows:

\begin{itemize}
\item choose $q$ on the range $(-0.5<q<+0.5)$ in a uniform random way;
\item choose $S_{1}^{*}=\pm1$ in a uniform random way, and then set $S_{2}^{*}=-S_{1}^{*}$;
\item choose $S_{0}=\pm1$ and $S_{0}^{*}=\pm1$ in uniform random ways;
\item solve the four constraining equations for $(p,\, M^{*},\, m_{0}/m_{0}^{*},\,\Psi)$;
\item calculate $\alpha$ from (\ref{eq28C}) for each signature in turn
and choose the value that satisfies $0<\alpha<1$, if one exists,
and record it;
\item repeat a large number of times.
\end{itemize}

\subsection{Results}

\begin{figure}
\subfigure[]{\includegraphics[%
  scale=0.32,
  angle=-90]{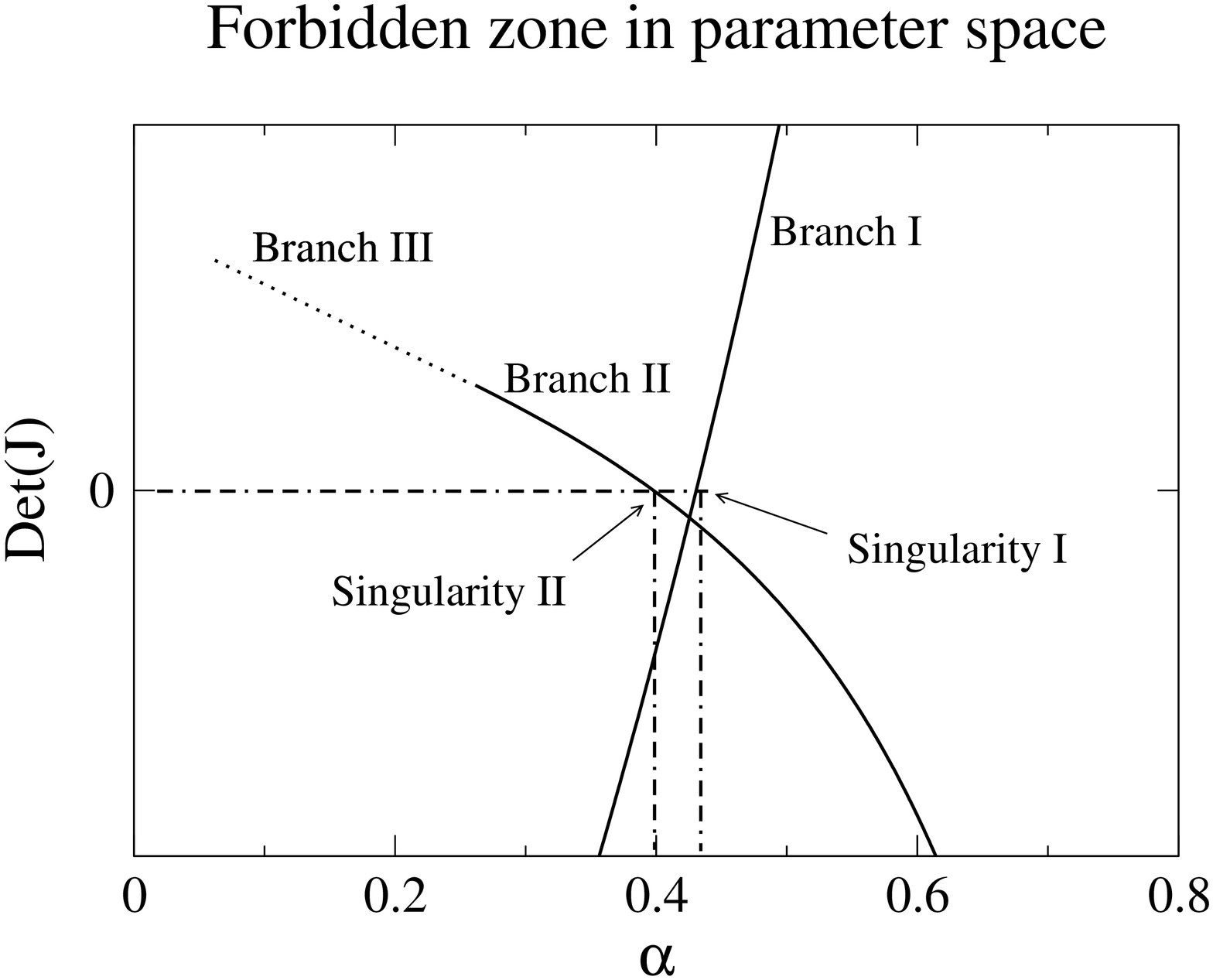}}\subfigure[]{\includegraphics[%
  scale=0.32,
  angle=-90]{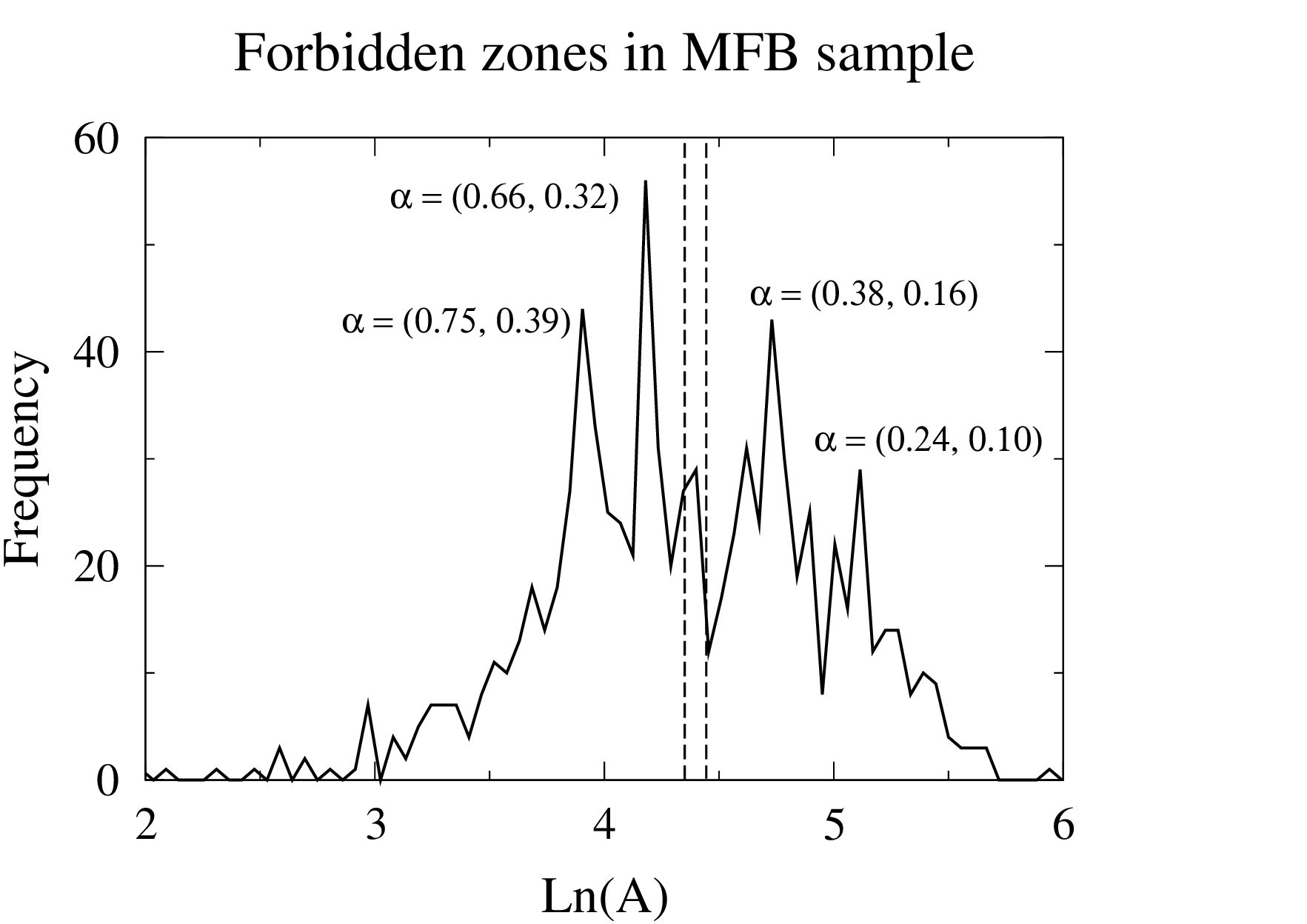}}

\caption{\label{Fig1} The left panel plots the determinant of the system
Jacobian against the parameter $\alpha$. It indicates the existance
of singularites at $\alpha=0.4$ and at $\alpha=0.43$. The right
panel, which primarily shows the $\ln A$ distribution for Mathewson
et al data, indicates the estimated positions of these two singularities
in $\ln A$ space. They clearly occupy a region of the parameter space
which is associated with an under-abundance of objects. }
\end{figure}
 Figure \ref{Fig1}a plots the determinant of the Jacobian of the
system against $\alpha$. It transpires that there are four distinct
branches of solutions, of which branch I and II are shown as solid
lines and branch III as a dotted line. Branch IV is out of range of
the diagram. The most immediate relevant features are: 

\begin{itemize}
\item on branch I, the Jacobian is singular at about $\alpha=0.43$;
\item on branch II, the Jacobian is singular at about $\alpha=0.40$.
\end{itemize}
Thus, if the theory does describe aspects of reality for disc galaxies,
then we would expect there to be a relative under-abundance of disc
galaxies in that part of the parameter space which is in the neigbourhood
of these singularities. 

However, there is a problem associated with considering $\alpha$-data
directly: whilst we know that $\alpha$ and $\ln A$ are extremely
strongly correlated (\cite{key-42,key-16}) we also know that estimates
of $\alpha$ from the data are an order of magnitude \emph{less} accurate
than those of $\ln A$ (Roscoe \cite{key-33}). Consequently, rather
than consider the $\alpha$ data directly, we consider it through
the proxy of $\ln A$. Figure \ref{Fig1}b shows what we find:

\begin{itemize}
\item The figure primarily shows the strong four-peak structure in the $\ln A$
distribution discussed in detail in Roscoe \cite{key-16} in the context
of the discrete dynamical states phenomenology;
\item Low values of $\ln A$ in the figure correspond to dim objects whilst
high values of $\ln A$ correspond to bright objects. Thus, there
are virtually no objects in the ranges $\ln A<3.5$ or $\ln A>5.5$;
\item The mean $\alpha$-values corresponding to each of the four major
peaks are indicated in the diagram;
\item The vertical dotted lines of figure \ref{Fig1}b indicate the estimated
$\ln A$ positions of the two singularities shown in $\alpha$-space
in figure \ref{Fig1}a;
\item We see that, whilst they occupy the middle of the $\ln A$ range where
we might expect to see the highest density of objects, they in fact
lie in a region where there is a very significant \emph{under-abundance}
of objects.
\end{itemize}
We can conclude that the theory's predictions about the existence
of singularities in the parameter space is in strong concordance with
that which is observed.

\section{Summary and conclusions}

The quasi-classical one-clock model has enjoyed a great deal of success
within the context of dynamics in the optical disc - but it fails
to admit the possibility of a transition to flatness within discs.
We then considered the possibility that a generalization of the quasi-classical
model into a quasi-relativistic form might provide a resolution to
this problem and found this to be the case. However, in order to merge
solutions between the quasi-classical and quasi-relativistic parts
of the disc, it then became necessary to consider jump-conditions
relating the two components of the disc. It transpired that the equations
representing these jump-conditions became degenerate in exactly that
part of the parameter space within which disc galaxies were already
know to be significantly under-abundant - that is, in the valley between
the two peaks corresponding to $\ln\approx4.17$ and $\ln A\approx4.72$
in figure \ref{Fig1}.

Although the valley between $4.17<\ln A<4.72$ is the deepest of the
valleys in figure \ref{Fig1}, it is not unique - if the theory is
able to provide an explanation for this deepest valley, we should
also expect it to provide an explanation for the valleys between $3.90<\ln A<4.17$
and $4.72<\ln A<5.10$ - either in explicit terms of why these regions
are underpopulated in discs or in explicit terms of why discs appear
to prefer the peaks. As things stand, the two-component disc model
does not do this. However, there are two major ways in which this
latter model is significantly over-simplified:

\begin{itemize}
\item Real disc galaxies have a central bulge region. In modelling terms,
this could be represented by an interior rotating spherical distribution
so that the over-all model would be a three-component disc model which
would have \emph{two} sets of jump-conditions placed upon the dynamics
- one from interior sphere to the middle disc and one from the middle
disc to the exterior flat disc. We might then expect these extra jump-conditions
to provide additional explanations for observed features;
\item Our model assumes perfect rotational symmetry within discs whereas
real discs have a very strong tendancy for a $180^{o}$ symmetry -
that is, to have two symmetrically opposed spiral arms. We might expect
the additional complexity arising from accounting for this two-arm
structure to give rise to significant improvement in the explanatory
power of the theory.
\end{itemize}
To conclude, the theory, which has the underlying structure of a LeSage-type
\emph{vacuum gravitation} theory, has `clocked up' sufficiently many
significant successes that one is inclined to believe that, at some
level, it is reflecting reality. This alone is enough to justify a
continued effort in its development. 

\appendix

\section{The possibility of transition to flatness within the quasi-classical
model\label{sec:The-possibility-of}}

It is sufficient to consider only equations (\ref{eq20}) and (\ref{eq21})
for which the parameters $(p,q)$ are independent and, once chosen,
fix the space of possible solutions. Suppose we consider the general
power-law solution which necessarily arises whenever $\rho=k/R^{2}$.
In this case, (\ref{eq20}) gives \begin{equation}
\epsilon^{2}=1,\,\,\,\, p\Psi^{2}+(1+p^{2}+q^{2})\Psi+p=0\label{eq21A}\end{equation}
so that, for any choice of $(p,q)$ then $\Psi$ is defined to have
one of two possible values, both constant. Thus, (\ref{eq21}) becomes
the simple power-law equation\[
\frac{1}{V}\frac{dV}{dR}=\frac{\alpha}{R}\]
for some constant $\alpha\equiv\alpha(p,q)$ which gives the exponent
in the solution, $V=AR^{\alpha}$ and, in Roscoe \cite{key-16}, we
have seen that it is this class of solutions which defines the mean
behaviour of optical rotation curves. 

Now, clearly, a flat rotation curve is a special case of the power-law
for which $\alpha=0$. Thus, in this special case we have the additional
relation\begin{equation}
\alpha\equiv1\mp\frac{p}{q}\mp\left(\frac{p^{2}+q^{2}}{q}\right)\frac{1}{\Psi}=0\label{eq21B}\end{equation}
where the $\mp$ arises because $\epsilon^{2}=1$. Thus, (\ref{eq21A})
and (\ref{eq21B}) together define a relationship $f(p,q)=0$ between
the parameters $(p,q)$ which must hold if $\alpha=0$. It follows
that if the conditions for transition to flatness in the radio disc
are satisfied within any given disc (that is, $f(p,q)=0$), then the
conditions for a \emph{non-zero} power-law solution within the optical
disc are excluded. But this is contrary to the conclusions the analyzes
of Roscoe \cite{key-42,key-16} for the generality of optical discs
- that is, there are substantial reasons for believing that the transition
to flatness does not occur within the context of the quasi-classical
model.

\section{Physical time and mathematical closure\label{sec:Temporal closure}}

The omitted details for the analysis of \S\ref{sub:Temporal closure}
are given below. Using the line element (\ref{eq1B}) we find\begin{equation}
ds^{2}\equiv d\mathbf{s}\cdot d\mathbf{s},\,\,\,\, d\mathbf{s}=S_{1}\sqrt{A}\, dR\,\mathbf{\hat{R}}+S_{2}\sqrt{B}\, R\, d\theta\,\hat{\theta}+S_{3}\sqrt{E}\, dT\,\mathbf{\hat{R}}\label{eq2AA}\end{equation}
subject to the condition that ${\cal M}$ must satisfy\begin{equation}
D=2S_{1}S_{3}\sqrt{AE}\rightarrow\left(\frac{\partial^{2}{\cal M}}{\partial R\partial T}\right)^{2}=\frac{\partial^{2}{\cal M}}{\partial R^{2}}\,\frac{\partial^{2}{\cal M}}{\partial T^{2}}.\label{eq4AA}\end{equation}
We can now use (\ref{eq2AA}) to define a generalized velocity in
the space\[
\mathbf{V}=\left(S_{1}\sqrt{A}\,\dot{R}+S_{3}\sqrt{E}\,\dot{T}\right)\mathbf{\hat{R}}+S_{2}\sqrt{B}\, R\,\dot{\theta}\,\hat{\theta}\]
from which a generalized acceleration vector, $\mathbf{\dot{V}}$,
can be defined as\begin{eqnarray*}
\mathbf{\dot{V}} & = & \left\{ S_{1}\frac{d}{d\tau}\left(\sqrt{A}\,\dot{R}\right)+S_{3}\frac{d}{d\tau}\left(\sqrt{E}\,\dot{T}\right)-S_{2}\sqrt{B}\, R\,\dot{\theta}^{2}\right\} \mathbf{\,\hat{R}}\\
\\ & + & \left\{ S_{2}\frac{d}{d\tau}\left(\sqrt{B}R\dot{\theta}\right)+S_{1}\sqrt{A}\,\dot{R}\,\dot{\theta}+S_{3}\sqrt{E}\,\dot{T}\,\dot{\theta}\right\} \,\hat{\theta}.\end{eqnarray*}
Using the condition that the ratio of radial to transverse accelerations
is to be constant within the disc, then, after multiplying through
by $S_{2}$, this latter equation leads to\begin{eqnarray}
\left\{ S_{1}S_{2}\frac{d}{d\tau}\left(\sqrt{A}\,\dot{R}\right)+S_{2}S_{3}\frac{d}{d\tau}\left(\sqrt{E}\,\dot{T}\right)-\sqrt{B}\, R\,\dot{\theta}^{2}\right\}  & =\nonumber \\
K\,\left\{ \frac{d}{d\tau}\left(\sqrt{B}R\dot{\theta}\right)+S_{1}S_{2}\sqrt{A}\,\dot{R}\,\dot{\theta}+S_{2}S_{3}\sqrt{E}\,\dot{T}\,\dot{\theta}\right\} \label{eq4A}\end{eqnarray}
as the generalization of (\ref{eq1C}). Since, from the earlier analysis,
we have that $A<0$, it follows from (\ref{eq4AA}) that $E<0$ also.
Since $B>0$ necessarily, then (\ref{eq4A}) is also complex so that
we must assume that $K$ is complex, as before. Adopting the conventions
that\[
\sqrt{A}=+j\sqrt{-A},\,\,\,\sqrt{E}=+j\sqrt{-E},\,\,\, S_{1}S_{2}\rightarrow S_{0},\,\,\, S_{2}S_{3}\rightarrow S_{1}\]
and defining\[
V_{\theta}\equiv R\dot{\theta},\,\, V_{R}\equiv\dot{R},\,\,\, K=q+pj\]
then (\ref{eq4A}) gives\begin{eqnarray*}
\frac{d}{d\tau}\left(\sqrt{B}V_{\theta}\right) & = & -q\sqrt{B}\frac{V_{\theta}^{2}}{R}-p\frac{d}{d\tau}\left\{ S_{0}\sqrt{-A}V_{R}+S_{1}\sqrt{-E}\dot{T}\right\} \\
\\\frac{V_{\theta}}{R}\left(S_{0}\sqrt{-A}V_{R}+S_{1}\sqrt{-E}\dot{T}\right) & = & -p\sqrt{B}\frac{V_{\theta}^{2}}{R}+q\frac{d}{d\tau}\left\{ S_{0}\sqrt{-A}V_{R}+S_{1}\sqrt{-E}\dot{T}\right\} \end{eqnarray*}
whilst (\ref{eq4AA}) becomes\[
D=-2S_{0}S_{1}\sqrt{-A}\sqrt{-E}\rightarrow\left(\frac{\partial^{2}{\cal M}}{\partial R\partial T}\right)^{2}=\frac{\partial^{2}{\cal M}}{\partial R^{2}}\,\frac{\partial^{2}{\cal M}}{\partial T^{2}}.\]

\section{A Rearrangement of the equations of motion\label{sec:A-Rearrangement-of}}

The key to a successful resolution of the system lies in recognizing
redundancy in a certain subset of the four basic equations of motion.
This subset is obtained by processing equations (\ref{eq5}), (\ref{eq7})
and (\ref{eq8}) in a certain way: From (\ref{eq5}) with use of (\ref{eq9})
we get directly\begin{eqnarray*}
V_{\theta}^{2}B\,(BR^{2}-m_{1}) & = & m_{1}\,\left(AV_{R}^{2}+DV_{R}\dot{T}+E\dot{T}^{2}\right)\\
\\ & = & m_{1}\,\left(S_{0}j\sqrt{-A}V_{R}+S_{1}j\sqrt{-E}\dot{T}\right)^{2}\\
 & \downarrow\\
V_{\theta}^{2}B\,\left(1-\frac{BR^{2}}{m_{1}}\right) & = & \left(S_{0}\sqrt{-A}V_{R}+S_{1}\sqrt{-E}\dot{T}\right)^{2}\end{eqnarray*}
Defining\[
\Psi\equiv S_{2}\sqrt{1-\frac{BR^{2}}{m_{1}}},\,\,\, S_{2}=\pm1\]
then we get\begin{equation}
V_{\theta}\sqrt{B}\Psi=S_{0}\sqrt{-A}V_{R}+S_{1}\sqrt{-E}\dot{T}.\label{eq10}\end{equation}
We can now write equations (\ref{eq7}) and (\ref{eq8}) as\begin{eqnarray*}
\frac{d}{d\tau}\left(\sqrt{B}V_{\theta}\right) & = & -q\sqrt{B}\frac{V_{\theta}^{2}}{R}-p\frac{d}{d\tau}\left(V_{\theta}\sqrt{B}\Psi\right)\\
\\\frac{V_{\theta}}{R}\left(V_{\theta}\sqrt{B}\Psi\right) & = & -p\sqrt{B}\frac{V_{\theta}^{2}}{R}+q\frac{d}{d\tau}\left(V_{\theta}\sqrt{B}\Psi\right)\end{eqnarray*}
We now eliminate $d(\sqrt{B}V_{\theta})/d\tau$ between these equations
to get, after some work,\[
\left\{ q+\frac{(1+p\Psi)(p+\Psi)}{q\Psi}\right\} V_{\theta}=\frac{\dot{\Psi}}{\Psi}R.\]
Finally, using the identity\[
\dot{\Psi}\equiv\frac{d\Psi}{d\tau}\equiv\frac{\partial\Psi}{\partial R}V_{R}+\frac{\partial\Psi}{\partial T}\dot{T}\]
(there is no $\theta$ dependency) we find\[
\left\{ q+\frac{(1+p\Psi)(p+\Psi)}{q\Psi}\right\} V_{\theta}-\frac{R}{\Psi}\frac{\partial\Psi}{\partial R}V_{R}=\frac{R}{\Psi}\frac{\partial\Psi}{\partial T}\dot{T}\]
Noting that (\ref{eq10}) and this latter equation are homogeneous
algebraic relations between $V_{\theta},\, V_{R}$ and $\dot{T}$
and that the ordinary transverse and radial velocity components are
given $v_{\theta}=V_{\theta}/\dot{T}$ and $v_{R}=V_{R}/\dot{T}$,
then we can write them as\begin{eqnarray*}
\sqrt{B}\Psi v_{\theta}-S_{0}\sqrt{-A}v_{R} & = & S_{1}\sqrt{-E}\\
\\\left\{ q+\frac{(1+p\Psi)(p+\Psi)}{q\Psi}\right\} v_{\theta}-\frac{R}{\Psi}\frac{\partial\Psi}{\partial R}v_{R} & = & \frac{R}{\Psi}\frac{\partial\Psi}{\partial T}\end{eqnarray*}
The structure of these two equations is fundamental to the system.

\section{The Variables Separable Class\label{sec:The-Variables-Separable}}

The variables-separable class solution of equation (\ref{eq9}) has
the general structure

\[
{\cal M}(R,T)=m_{0}(R+\beta_{0})^{\mu}(T+\beta_{1})^{1-\mu}+\beta_{2}\]
where $(m_{0},\beta_{0},\beta_{1},\beta_{2})$ are constants.

We ask under what conditions (if any) this solution (with $\beta_{0}=\beta_{1}=0$
for simplicity) can satisfy condition (\ref{eq13}). Using the definitions
(\ref{eq1D}) and (\ref{eq9A}) we easily find that the variables
separable class can only be a solution for the particular case ${\cal M}(R,T)=m_{0}\sqrt{RT}$,
and $S_{0}=S_{1}=\pm1$. If these conditions are satisfied, then we
find that both of (\ref{eq14}) reduce to a single algebraic condition
having the general structure\[
f(RT,M)=0,\,\,\,\,{\rm where}\,\,\, M\equiv\frac{m_{0}}{m_{1}}\]
and $m_{1}$ is the mass-scaling constant in the function $\Psi$
defined at (\ref{eq9A}). But $f(RT,M)=0$ implies directly that $RT=const$
so that, finally, we find that the variables separable class is a
solution subject to the conditions\[
{\cal M}(R,T)=\sqrt{RT},\,\,\, RT=const,\,\,\, S_{0}S_{1}=+1,\,\,\, f(1,M)=0\,\,\,{\rm where}\,\, M\equiv\frac{m_{0}}{m_{1}}.\]
This corresponds to ${\cal M}=const$ on the characteristic surface
$R=const/T$. As we shall show, it transpires that this class of solution
is associated with a degeneracy in the system which removes them from
the current discussion.

\section{Expansion of an Equation of Motion\label{sec:Expansion-of-an}}

Equation (\ref{eq16}) is given by

\begin{equation}
\frac{d}{d\tau}\left\{ \frac{1}{2{\cal L}}\left(D\dot{R}+2E\dot{T}\right)\right\} -\frac{\partial{\cal L}}{\partial T}=0\label{eq22}\end{equation}
where\begin{eqnarray*}
{\cal L} & \equiv & \left(A\dot{R}^{2}+BR^{2}\dot{\theta}^{2}+D\dot{R}\dot{T}+E\dot{T}^{2}\right)^{1/2}=\dot{T}\left(Av_{R}^{2}+Bv_{\theta}^{2}+Dv_{R}+E\right)^{1/2}\equiv\dot{T}L\end{eqnarray*}
and $A,\, B,\, D$ and $E$ are defined at (\ref{eq1D}). Thus (\ref{eq22})
can be written as\begin{equation}
\frac{d}{dT}\left\{ \frac{1}{2L}\left(Dv_{R}+2E\right)\right\} -\frac{1}{2L}\left(\frac{\partial A}{\partial T}v_{R}^{2}+\frac{\partial B}{\partial T}v_{\theta}^{2}+\frac{\partial D}{\partial T}v_{R}+\frac{\partial E}{\partial T}\right)\label{eq23}\end{equation}
where we have used the identity\[
\frac{1}{\dot{T}}\frac{d}{d\tau}\equiv\frac{d}{dT}.\]
Equations (\ref{eq5}) and (\ref{eq11}) give, respectively\[
L=\frac{1}{\sqrt{m_{1}}}RBv_{\theta},\,\,\,\,\, v_{R}=S_{0}\frac{\sqrt{B}}{\sqrt{-A}}\Psi v_{\theta}-S_{0}S_{1}\frac{\sqrt{-E}}{\sqrt{-A}}.\]
Substituting these in (\ref{eq23}), we find that the velocity derivatives,
which all arise from the term,\[
\frac{d}{dT}\left\{ \frac{Dv_{R}}{2L}\right\} \]
vanish identically so that (\ref{eq23}) becomes purely algebraic
in the velocity terms: in terms of $L,\, v_{R}$ and $v_{\theta}$
it becomes\begin{eqnarray*}
\frac{\partial A}{\partial T}v_{R}^{2}+\frac{\partial B}{\partial T}v_{\theta}^{2}-S_{0}D\frac{d}{dT}\left(\frac{\Psi\sqrt{B}}{\sqrt{-A}}\right)v_{\theta}+\frac{\partial E}{\partial T}-2\frac{dE}{dT}\\
+S_{0}S_{1}D\frac{d}{dT}\left(\frac{\sqrt{-E}}{\sqrt{-A}}\right)+\frac{(Dv_{R}+2E)}{RB}\frac{d}{dT}\left(RB\right) & = & 0.\end{eqnarray*}
using the definitions of $A,\, B,\, D$ and $E$ given at (\ref{eq1D}),
and using Maple, we obtain the remarkable result that\[
v_{\theta}=\pm\frac{1}{\sqrt{2-M}}\]
on the characteristic $R=1+T$ where $M\equiv m_{0}/m_{1}.$

\section{Conservation of Generalized Angular Momentum\label{sec:Conservation-of-Generalized}}

In the following, we give the details omitted in \S\ref{sub:Conservation-of-Generalized}.
Reference to (\ref{eq5}) shows that the quantity $R^{2}\dot{\theta}B/{\cal L}$
must be conserved across the transition boundary. Using the no-slip
condition $v_{\theta}^{*}=v_{\theta}$ at $R=R_{T}$, this requirement
reduces to\[
\frac{B^{*}}{{\cal L}^{*}}=\frac{B}{{\cal L}}\]
at the boundary so that, using (\ref{eq1E}) for ${\cal L}$ and the
corresponding definition for ${\cal L}^{*}$ we get\[
\frac{B^{*}}{\sqrt{A^{*}v_{R}^{*2}+B^{*}v_{\theta}^{*2}}}=\frac{B}{\sqrt{Av_{R}^{2}+Bv_{\theta}^{2}+Dv_{R}+E}}.\]
Using (\ref{eq9}), (\ref{eq24}) and (\ref{eq25}) this latter condition
becomes\[
\frac{\sqrt{m_{0}^{*}}}{\sqrt{v_{\theta}^{*2}-v_{R}^{*2}}}=\frac{\sqrt{m_{0}}}{\sqrt{-v_{R}^{2}+2v_{\theta}^{2}-2S_{0}S_{1}v_{R}-1}}\]
We now consider the left-hand side and right-hand sides separately:

\begin{description}
\item [Left-hand~side]~
\end{description}
Using (\ref{eq25A}) and (\ref{eq27}) we get\[
lhs\equiv\frac{\sqrt{m_{0}^{*}}}{\sqrt{v_{\theta}^{*2}-v_{R}^{*2}}}=\frac{\sqrt{m_{0}^{*}}}{v_{\theta}^{*}\sqrt{1-\Psi^{*2}}}=\frac{\sqrt{m_{0}^{*}}}{v_{\theta}^{*}}\frac{\sqrt{m_{1}^{*}}}{\sqrt{m_{0}^{*}}}=\frac{\sqrt{m_{1}^{*}}}{v_{\theta}^{*}}=\sqrt{m_{1}^{*}}\sqrt{2-M}\]
where the no-slip condition, $v_{\theta}^{*}=v_{\theta}$, and (\ref{eq18A})
have been used for the last step.

\begin{description}
\item [Right-hand~side]~
\end{description}
Equations (\ref{eq25B}) and (\ref{eq18A}) together give the scaled
radial velocity in the hyperbolic disc as:\[
v_{R}=S_{0}(S_{2}-S_{1}).\]
Using (\ref{eq27B}), we deduce immediately that $v_{R}=-2S_{0}S_{1}$.
Using this and (\ref{eq18A}) we get\[
rhs\equiv\frac{\sqrt{m_{0}}}{\sqrt{-v_{R}^{2}+2v_{\theta}^{2}-2S_{0}S_{1}v_{R}-1}}=\sqrt{m_{0}}\frac{\sqrt{2-M}}{\sqrt{M}}=\sqrt{m_{1}}\sqrt{2-M}.\]

\begin{description}
\item [lhs=rhs]~
\end{description}
Putting these together, we finally get \[
m_{1}^{*}=m_{1}\]
as the jump-condition which guarantees the conservation of generalized
angular momentum across the transition boundary.

\section{Conservation of mass flow\label{sec:Conservation-of-mass}}

In the following, we give the details omitted in \S\ref{sub:Conservation-of-mass}.
If we use $R=R_{T}$ to denote the radial transition boundary, the
mass-flow conservation across this boundary can be stated as:\[
\left.\rho^{*}v_{R}^{*}\right|_{R=R_{T}}=\left.\rho v_{R}\right|_{R=R_{T}}.\]
Since \[
B\equiv\frac{1}{R}\frac{\partial{\cal M}}{\partial R}=2\pi\rho\]
on both parts of the disc, then we can write the mass-flow condition
as\begin{equation}
\left.B^{*}v_{R}^{*}\right|_{R=R_{T}}=\left.Bv_{R}\right|_{R=R_{T}}.\label{eq26}\end{equation}
Thus, we need appropriate expressions for $(B^{*},v_{R}^{*})$ and
$(B,v_{R})$. From \S\ref{sub:The-power-law-as} we have already
noted that ${\cal M}^{*}=m_{0}^{*}\ln R$ so that using the definitions
given at (\ref{eq1D}), then\begin{equation}
A^{*}=-\frac{m_{0}^{*}}{R^{2}},\,\,\, B^{*}=\frac{m_{0}^{*}}{R^{2}}\label{eq24}\end{equation}
whilst in \S\ref{sub:The-Mass-Equation} we have noted that ${\cal M}=m_{0}\ln(1+R+T)$
on $R=1+T$ so that using the definitions given at (\ref{eq1D}),
then\begin{equation}
A=-\frac{m_{0}}{4R^{2}},\,\,\, B=\frac{m_{0}}{2R^{2}},\,\,\, E=-\frac{m_{0}}{4R^{2}}\label{eq25}\end{equation}
also on $R=1+T$. Noting that the quasi-classical relations can be
obtained from the quasi-relativistic relations simply by dropping
of all time-dependent terms then (\ref{eq15}) gives, after using
(\ref{eq24}) and (\ref{eq9A}), \begin{equation}
v_{R}^{*}=S_{0}^{*}\frac{\sqrt{B^{*}}}{\sqrt{-A^{*}}}\Psi^{*}v_{\theta}^{*}=S_{0}^{*}\Psi^{*}v_{\theta}^{*}\label{eq25A}\end{equation}
whilst use of (\ref{eq15}) directly gives, after using (\ref{eq25})
and (\ref{eq9A}), \begin{equation}
v_{R}=S_{0}\frac{\sqrt{B}}{\sqrt{-A}}\Psi v_{\theta}-S_{0}S_{1}\frac{\sqrt{-E}}{\sqrt{-A}}=S_{0}S_{2}\sqrt{2-M}v_{\theta}-S_{0}S_{1}.\label{eq25B}\end{equation}
Hence, with these definitions of $B$, $B^{*}$, $v_{R}$, $v_{R}^{*}$
and the no-slip condition $v_{\theta}=v_{\theta}^{*}$ at transition,
then (\ref{eq26}) can be written as\begin{eqnarray*}
\frac{m_{0}^{*}}{R^{2}}S_{0}^{*}\Psi^{*}v_{\theta} & = & \frac{m_{0}}{2R^{2}}\left(S_{0}S_{2}\sqrt{2-M}v_{\theta}-S_{0}S_{1}\right)\\
 & \downarrow\\
\Psi^{*} & = & \frac{m_{0}}{2m_{0}^{*}}S_{0}^{*}S_{0}\left(S_{2}\sqrt{2-M}-\frac{S_{1}}{v_{\theta}}\right)\end{eqnarray*}
If we now use the flatness solution given at (\ref{eq18A}) we obtain
our jump-condition for mass conservation as\[
\Psi^{*}=\frac{m_{0}}{m_{0}^{*}}S_{0}^{*}S_{0}\frac{\sqrt{2-M}}{2}\left(S_{2}-S_{1}\right),\,\,\, M\equiv\frac{m_{0}}{m_{1}}\]
where also, given the definition at (\ref{eq9A}),\[
\Psi^{*}\equiv S_{2}^{*}\sqrt{1-M^{*}},\,\,\, M^{*}\equiv\frac{m_{0}^{*}}{m_{1}^{*}}.\]

\end{document}